\documentclass{llncs}
\usepackage{graphicx,xspace}

\newcommand{\Oh}[1]
	{\ensuremath{\mathcal{O}\!\left({#1}\right)}\xspace}
\newcommand{\IN}
	{\ensuremath{\mathtt{IN}}\xspace}
\newcommand{\OUT}
	{\ensuremath{\mathtt{OUT}}\xspace}
\newcommand{\TRUE}
  {\ensuremath{\mathtt{1}}\xspace}
\newcommand{\FALSE}
  {\ensuremath{\mathtt{0}}\xspace}
\newcommand*{\ie}{i.e.,\@\xspace}

\begin{document}

\title{Fully Dynamic de Bruijn Graphs}
\author{Djamal Belazzougui\inst{1} \and
  Travis Gagie\inst{2,3} \and
  Veli M\"akinen\inst{2,3} \and
  Marco Previtali\inst{4}} 
\institute{CERIST, Algeria \and
  Helsinki Institute for Information Technology, Finland \and
  University of Helsinki, Finland \and
  University of Milano-Bicocca, Italy}
\maketitle

\begin{abstract}
We present a space- and time-efficient fully dynamic implementation de Bruijn graphs, which can also support fixed-length jumbled pattern matching.
\end{abstract}

\section{Introduction}
\label{sec:introduction}
Bioinformaticians define the $k$th-order de Bruijn graph for a string or set of strings to be the directed graph whose nodes are the distinct $k$-tuples in those strings and in which there is an edge from $u$ to $v$ if there is a \((k + 1)\)-tuple somewhere in those strings whose prefix of length $k$ is $u$ and whose suffix of length $k$ is $v$.\footnote{An alternative definition, which our data structure can be made to handle but which we do not consider in this paper, has an edge from $u$ to $v$ whenever both nodes are in the graph.}  These graphs have many uses in bioinformatics, including {\it de novo\/} assembly~\cite{zerbino2008velvet}, read correction~\cite{DBLP:journals/bioinformatics/SalmelaR14} and pan-genomics~\cite{siren2014indexing}.  The datasets in these applications are massive and the graphs can be even larger, however, so pointer-based implementations are impractical.  Researchers have suggested several approaches to representing de Bruijn graphs compactly, the two most popular of which are based on Bloom filters~\cite{wabi,cascading} and the Burrows-Wheeler Transform~\cite{bowe2012succinct,boucher2015variable,belazzougui2016bidirectional}, respectively.  In this paper we describe a new approach, based on minimal perfect hash functions~\cite{mehlhorn1982program}, that is similar to that using Bloom filters but has better theoretical bounds when the number of connected components in the graph is small, and is fully dynamic: i.e., we can both insert and delete nodes and edges efficiently, whereas implementations based on Bloom filters are usually semi-dynamic and support only insertions.  We also show how to modify our implementation to support, e.g., jumbled pattern matching~\cite{BCFL12} with fixed-length patterns.

Our data structure is based on a combination of Karp-Rabin hashing~\cite{KR87} and minimal perfect hashing, which we will describe in the full version of this paper and which we summarize for now with the following technical lemmas:

\begin{lemma}
\label{lem:static}
Given a static set $N$ of $n$ $k$-tuples over an alphabet $\Sigma$ of size $\sigma$, with high probability in  $O(kn)$ expected time we can build a function \(f : \Sigma^k \rightarrow \{0, \ldots, n - 1\}\) with the following properties:
\begin{itemize}
\item when its domain is restricted to $N$, $f$ is bijective;
\item we can store $f$ in $O(n + \log k+\log\sigma)$ bits;
\item given a $k$-tuple $v$, we can compute \(f (v)\) in $\Oh{k}$ time;
\item given $u$ and $v$ such that the suffix of $u$ of length \(k - 1\) is the prefix of $v$ of length \(k - 1\), or vice versa, if we have already computed \(f (u)\) then we can compute \(f (v)\) in $\Oh{1}$ time.
\end{itemize}
\end{lemma}

\begin{lemma}
\label{lem:dynamic}
If $N$ is dynamic then we can maintain a function $f$ as described in Lemma~\ref{lem:static} except that:\
\begin{itemize}
\item the range of $f$ becomes \(\{0, \ldots, 3 n - 1\}\);
\item when its domain is restricted to $N$, $f$ is injective;
\item our space bound for $f$ is $\Oh{n (\log \log n + \log \log \sigma)}$ bits with high probability;
\item insertions and deletions take $\Oh{k}$ amortized expected time.
\item the data structure may work incorrectly with very low probability
(inversely polynomial in $n$). 
\end{itemize}
\end{lemma}

Suppose $N$ is the node-set of a de Bruijn graph.  In Section~\ref{sec:static} we show how we can store $\Oh{n \sigma}$ more bits than Lemma~\ref{lem:static} such that, given a pair of $k$-tuples $u$ and $v$ of which at least one is in $N$, we can check whether the edge \((u, v)\) is in the graph.  This means that, if we start with a $k$-tuple in $N$, then we can explore the entire connected component containing that $k$-tuple in the underlying undirected graph.  On the other hand, if we start with a $k$-tuple not in $N$, then we will learn that fact as soon as we try to cross an edge to a $k$-tuple that is in $N$.  To deal with the possibility that we never try to cross such an edge, however --- \ie that our encoding as described so far is consistent with a graph containing a connected component disjoint from $N$ --- we cover the vertices with a forest of shallow rooted trees.  We store each root as a $k$-tuple, and for each other node we store \(1 + \lg \sigma\) bits indicating which of its incident edges leads to its parent.  To verify that a $k$-tuple we are considering is indeed in the graph, we ascend to the root of the tree that contains it and check that $k$-tuple is what we expect.  The main challenge for making our representation dynamic with Lemma~\ref{lem:dynamic} is updating the covering forest.  In Section~\ref{sec:dynamic} how we can do this efficiently while maintaining our depth and size invariants.  Finally, in Section~\ref{sec:jumbled} we observe that our representation can be easily modified for other applications by replacing the Karp-Rabin hash function by other kinds of hash functions.  To support jumbled pattern matching with fixed-length patterns, for example, we hash the histograms indicating the characters' frequencies in the $k$-tuples.

\section{Static de Bruijn Graphs}
\label{sec:static}

Let \(G\) be a de Bruijn graph of order \(k\), let \(N = \{v_0, \ldots, v_{n-1}\}\) be the set of its nodes, and let \(E = \{a_0, \ldots, a_{e-1}\}\) be the set of its edges. We call each \(v_i\) either a node or a \(k\)-tuple, using interchangeably the two terms since there is a one-to-one correspondence between nodes and labels.

We maintain the structure of \(G\) by storing two binary matrices, \IN and \OUT, of size \(n \times \sigma\). For each node, the former represents its incoming edges whereas the latter represents its outgoing edges. In particular, for each \(k\)-tuple \(v_x = c_1 c_2 \ldots c_{k-1} a\), the former stores a row of length \(\sigma\) such that, if there exists another \(k\)-tuple \(v_y = b c_1 c_2 \ldots c_{k-1}\) and an edge from \(v_y\) to \(v_x\), then the position indexed by \(b\) of such row is set to \TRUE. Similarly, \OUT contains a row for \(v_y\) and the position indexed by \(a\) is set to \TRUE. As previously stated, each \(k\)-tuple is uniquely mapped to a value between \(0\) and \(n-1\) by \(f\), where $f$ is as defined in Lemma~\ref{lem:static}, and therefore we can use these values as indices for the rows of the matrices \IN and \OUT, \ie in the previous example the values of \(\IN[f(v_x)][b]\) and \(\OUT[f(v_y)][a]\) are set to \TRUE.  We note that, e.g., the SPAdes assembler~\cite{Ban12} also uses such matrices.

Suppose we want to check whether there is an edge from \(b X\) to \(X a\).
Letting \(f(b X) = i\) and \(f(X a) = j\), we first assume \(b X\) is in \(G\) and check the values of \(\OUT [i] [a] \) and \( \IN [j] [b]\). If both values are \TRUE, we report that the edge is present and we say that the edge is \emph{confirmed} by \IN and \OUT; otherwise, if any of the two values is \FALSE, we report that the edge is absent. Moreover, note that if \(b X\) is in \(G\) and \(\OUT [i] [a] = \TRUE\), then \(X a\) is in \(G\) as well. Symmetrically, if \(X a\) is in \(G\) and \(\IN [j] [b] = \TRUE\), then \(b X\) is in \(G\) as well. Therefore, if \(\OUT [i] [a] = \IN [j] [b] = \TRUE\), then \(b X\) is in \(G\) if and only if \(X a\) is. This means that, if we have a path \(P\) and if all the edges in \(P\) are confirmed by \IN and \OUT, then either all the nodes touched by \(P\) are in \(G\) or none of them is.

\begin{figure}[t!]
\begin{center}
\includegraphics[width=\textwidth]{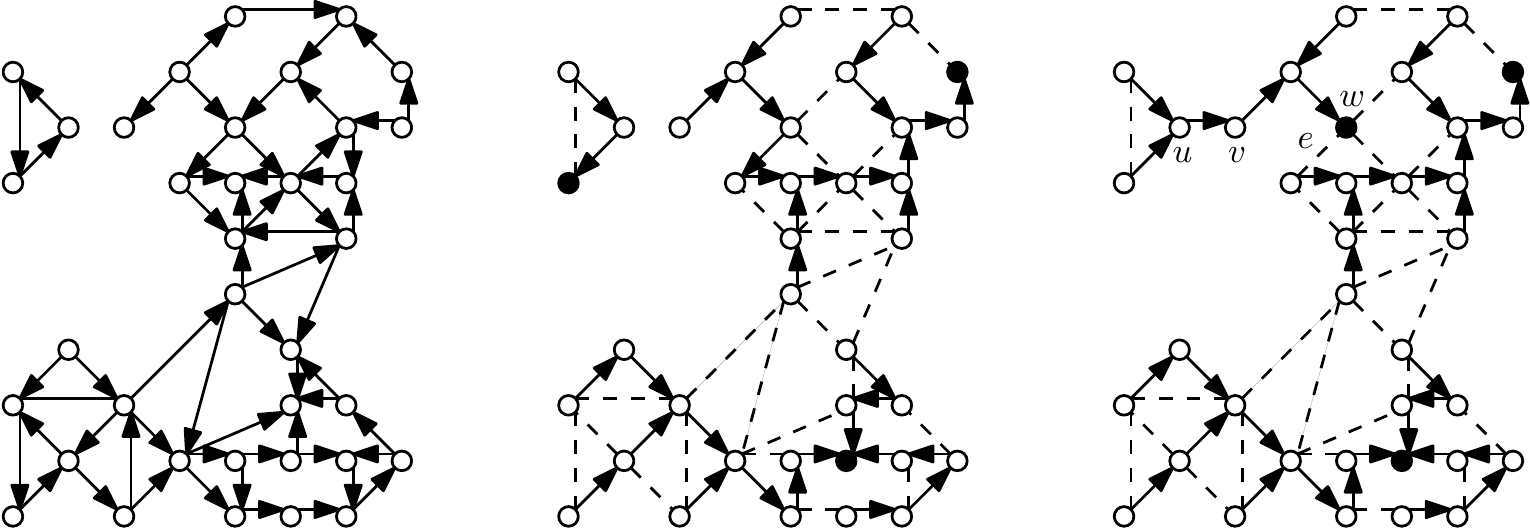}
\caption{Given a de Bruijn graph (left), we cover the underlying undirected graph with a forest of rooted trees of height at most \(3 k \lg \sigma\) (center).  The roots are shown as filled nodes, and parent pointers are shown as arrows; notice that the directions of the arrows in our forest are not related to the edges' directions in the original de Bruijn graph.  We sample the $k$-tuples at the roots so that, starting at a node we think is in the graph, we can verify its presence by finding the root of its tree and checking its label in $\Oh{k \log \sigma}$ time.  The most complicated kind of update (right) is adding an edge between a node $u$ in a small connected component to a node $v$ in a large one, $v$'s depth is more than \(2 k \lg \sigma\) in its tree.  We re-orient the parent pointers in $u$'s tree to make $u$ the temporary root, then make $u$ point to $v$.  We ascend \(k \lg \sigma\) steps from $v$, then delete the parent pointer $e$ of the node $w$ we reach, making $w$ a new root.  (To keep this figure reasonably small, some distances in this example are smaller than prescribed by our formulas.)}
\label{fig:trees}
\end{center}
\end{figure}

We now focus on detecting false positives in our data structure maintaining a reasonable memory usage. Our strategy is to sample a subset of nodes for which we store the plain-text \(k\)-tuple and connect all the unsampled nodes to the sampled ones. More precisely, we partition nodes in the undirected graph \(G^\prime\) underlying \(G\) into a forest of rooted trees of height at least \(k \lg \sigma \) and at most \(3 k \lg \sigma\). For each node we store a pointer to its parent in the tree, which takes \(1 + \lg \sigma\) bits per node, and we sample the \(k\)-mer at the root of such tree. We allow a tree to have height smaller than \(k \lg \sigma\) when necessary, 
e.g., if it covers a connected component.  Figure~\ref{fig:trees} shows an illustration of this idea.

We can therefore check whether a given node \(v_x\) is in \(G\) by first computing \(f(v_x)\) and then checking and ascending at most \(3 k \lg \sigma\) edges, updating \(v_x\) and \(f(v_x)\) as we go. Once we reach the root of the tree we can compare the resulting \(k\)-tuple with the one sampled to check if \(v_x\) is in the graph. This procedure requires \Oh{k \lg \sigma} time since computing the first value of \(f(v_x)\) requires \Oh{k}, ascending the tree requires constant time per edge, and comparing the \(k\)-tuples requires \Oh{k}.

We now describe a Las Vegas algorithm for the construction of this data structure that requires, with high probability, \Oh{kn + n\sigma} expected time.
We recall that \(N\) is the set of input nodes of size \(n\).
We first select a function \(f\) and construct bitvector \(B\) of size \(n\) initialized with all its elements set to \FALSE.
For each elements \(v_x\) of \(N\) we compute \(f(v_x) = i\) and check the value of \(B[i]\).
If this value is \FALSE we set it to \TRUE and proceed with the next element in \(N\), if it is already set to \TRUE, we reset \(B\), select a different function \(f\), and restart the procedure from the first element in \(N\).
Once we finish this procedure --- \ie we found that \(f\) do not produces collisions when applied to \(N\) --- we store \(f\) and proceed to initialize \IN and \OUT correctly.
This procedure requires with high probability \Oh{kn} expected time for constructing \(f\) and \Oh{n\sigma} time for computing \IN and \OUT.
Notice that if \(N\) is the set of \(k\)-tuples of a single text sorted by their starting position in the text, each \(f(v_x)\) can be computed in constant time from \(f(v_{x-1})\) except for \(f(v_0)\) that still requires \Oh{k}.
More generally, if \(N\) is the set of \(k\)-tuples of \(t\) texts sorted by their initial position, we can compute \(n - t\) values of the function \(f(v_x)\) in constant time from \(f(v_{x-1})\) and the remaining in \Oh{k}.
We will explain how to build the forest in the full version of this paper.
In this case the construction requires, with high probability, \(\Oh{kt + n + n\sigma} = \Oh{kt + n\sigma}\) expected time.

Combining our forest with Lemma~\ref{lem:static}, we can summarize our static data structure in the following theorem:
\begin{theorem}
\label{thm:static}
Given a static $\sigma$-ary $k$th-order de Bruijn graph $G$ with $n$ nodes, with high probability in $\Oh{k n + n \sigma}$ 
expected time we can store $G$ in $\Oh{\sigma n}$ bits plus $\Oh{k \log \sigma}$ bits for each connected component in the
underlying undirected graph, such that checking whether a node is in $G$ takes $\Oh{k \log \sigma}$ time, listing the edges incident to a node we are visiting takes $\Oh{\sigma}$ time, and crossing an edge takes $\Oh{1}$ time.
\end{theorem}
In the full version we will show how to use monotone minimal perfect hashing~\cite{BBPV09}
to reduce the space to $(2+\epsilon)n\sigma$ bits of space
(for any constant $\epsilon>0$). We will also show how to reduce the time to list the 
edges incident to a node of degree $d$ to $O(d)$, and the time 
to check whether a node is in $G$ to $\Oh{k}$. We note that the obtained space and query times are 
both optimal up to constant factors, which is unlike previous methods which have additional 
factor(s) depending on $k$ and/or $\sigma$ in space and/or time. 
\section{Dynamic de Bruijn Graphs}
\label{sec:dynamic}

In the previous section we presented a static representation of de Buijn graphs, we now present how we can make this data structure dynamic. In particular, we will show how we can insert and remove edges and nodes and that updating the graph reduces to managing the covering forest over \(G\). 
In this section, when we refer to $f$ we mean the function defined in Lemma~\ref{lem:dynamic}.
We first show how to add or remove an edge in the graph and will later describe how to add or remove a node in it.
The updates must maintain the following invariant: any tree must have size at least $k\log\sigma$
and height at most $3k\log\sigma$ except when the tree covers (all nodes in) a connected component 
of size at most $k\log\sigma$. 

Let \(v_x\) and \(v_y\) be two nodes in \(G\), \(e = (v_x, v_y)\) be an edge in \(G\), and let \(f(v_x) = i\) and \(f(v_y) = j\).

Suppose we want to add \(e\) to \(G\). First, we set to \TRUE the values of \(\OUT[i][a]\) and \(\IN[j][b]\) in constant time.
We then check whether \(v_x\) or \(v_y\) are in different components of size less than \(k \lg \sigma\) in \Oh{k \lg \sigma} time for each node.
If both components have size greater than \(k \lg \sigma\) we do not have to proceed further since the trees will not change.  If both connected components have size less than \(k \lg \sigma\) we merge their trees in \Oh{k \lg \sigma} time by traversing both trees and switching the orientation of the edges in them, discarding the samples at the roots of the old trees and sampling the new root in \Oh{k} time.

If only one of the two connected components has size greater than \(k \lg \sigma\) we select it and perform a tree traversal to check whether the depth of the node is less than \(2 k \lg \sigma\).
If it is, we connect the two trees as in the previous case.  If it is not, we traverse the tree in the bigger components upwards for \(k \lg \sigma\) steps, we delete the edge pointing to the parent of the node we reached creating a new tree, and merge it with the smaller one.
This procedure requires \Oh{k \lg \sigma} time since deleting the edge pointing to the parent in the tree requires \Oh{1} time, \ie we have to reset the pointer to the parent in only one node.

Suppose now that we want to remove \(e\) from \(G\).
First we set to \FALSE the values of \(\OUT[i][a]\) and \(\IN[j][b]\) in constant time.
Then, we check in \Oh{k} time whether \(e\) is an edge in some tree by computing  \(f(v_x)\) and \(f(v_y)\) checking for each node if that edge is the one that points to their parent.
If \(e\) is not in any tree we do not have to proceed further whereas if it is we check the size of each tree in which \(v_x\) and \(v_y\) are.
If any of the two trees is small (\ie if it has fewer than \(k \lg \sigma\) elements) we search any outgoing edge from the tree that connects it to some other tree. 
If such an edge is not found we conclude that we are in a small connected component that is covered by the current tree and we sample a node in the tree as a root and switch directions of some edges if necessary. 
If such an edge is found, we merge the small tree with the bigger one by adding the edge and switch the direction of some edges originating from the small tree if necessary. Finally if the height of the new tree exceeds $3k\log\sigma$, we traverse the tree upwards from the deepest node in the tree (which was necessarily a node in the smaller tree before the merger) for \(2k \lg \sigma\) steps, delete the edge pointing to the parent of the reached node, creating a new tree. 
This procedure requires
$\Oh{k \lg \sigma}$ since the number of nodes traversed is at most \(O(k \lg \sigma)\)
and the number of changes to the data structures is also at most \(O(k \lg \sigma)\)
with each change taking expected constant time.



It is clear that the insertion and deletion algorithms will maintain the invariant on the tree sizes. 
It is also clear that the invariant implies that the number of sampled nodes is $O(n/(k\log\sigma))$ plus the 
number of connected components. 

We now show how to add and remove a node from the graph.
Adding a node is trivial since it will not have any edge connecting it to any other node.
Therefore adding a node reduces to modify the function \(f\) and requires \Oh{k} amortized expected time.
When we want to remove a node, we first remove all its edges one by one and, once the node is isolated from the graph, we remove it by updating the function \(f\).
Since a node will have at most \(\sigma\) edges and updating \(f\) requires \Oh{k} amortized expected time, the amortized expected time complexity of this procedure is
$\Oh{\sigma k\lg \sigma+ k}$.

Combining these techniques for updating our forest with Lemma~\ref{lem:dynamic}, we can summarize our dynamic data structure in the following theorem:

\begin{theorem}
\label{thm:dynamic}
We can maintain a $\sigma$-ary $k$th-order de Bruijn graph $G$ with $n$ nodes that is fully dynamic (i.e., supporting node and edge insertions and deletions) in $\Oh{n (\log \log n + \sigma)}$ bits (plus $\Oh{k \log \sigma}$ bits for each connected component) with high probability, such that we can add or remove an edge in expected \Oh{k\lg\sigma} time, add a node in expected \Oh{k+\sigma} time, and remove a node in expected \Oh{\sigma k\lg \sigma} time, and queries have the same time bounds as in Theorem~\ref{thm:static}. The data structure may work incorrectly with very low probability (inversely polynomial in $n$). 
\end{theorem}

\section{Jumbled Pattern Matching}
\label{sec:jumbled}

Karp-Rabin hash functions implicitly divide their domain into equivalence classes --- i.e., subsets in which the elements hash to the same value.  In this paper we have chosen Karp-Rabin hash functions such that each equivalence class contains only one $k$-tuple in the graph.  Most of our efforts have gone into being able, given a $k$-tuple and a hash value, to determine whether that $k$-tuple is the unique element of its equivalence class in the graph.  In some sense, therefore, we have treated the equivalence relation induced by our hash functions as a necessary evil, useful for space-efficiency but otherwise an obstacle to be overcome.  For some applications, however --- e.g., parameterized pattern matching, circular pattern matching or jumbled pattern matching --- we are given an interesting equivalence relation on strings and asked to preprocess a text such that later, given a pattern, we can determine whether any substrings of the text are in the same equivalence class as the pattern.  We can modify our data structure for some of these applications by replacing the Karp-Rabin hash function by other kinds of hash functions.

For indexed jumbled pattern matching~\cite{BCFL12,KRR13,ACLL14} we are asked to pre-process a text such that later, given a pattern, we can determine quickly whether any substring of the text consists of exactly the same multiset of characters in the pattern.  Consider fixed-length jumbled pattern matching, when the length of the patterns is fixed at pre-processing time.  If we modify Lemmas~\ref{lem:static} and~\ref{lem:dynamic} so that, instead of using Karp-Rabin hashes in the definition of the function $f$, we use a hash function on the histograms of characters' frequencies in $k$-tuples, our function $f$ will map all permutations of a $k$-tuple to the same value.  The rest of our implementation stays the same, but now the nodes of our graph are multisets of characters of size $k$ and there is an edge between two nodes $u$ and $v$ if it is possible to replace an element of $u$ and obtain $v$.  If we build our graph for the multisets of characters in $k$-tuples in a string $S$, then our process for checking whether a node is in the graph tells us whether there is a jumbled match in $S$ for a pattern of length $k$.  If we build a tree in which the root is a graph for all of $S$, the left and right children of the root are graphs for the first and second halves of $S$, etc., as described by Gagie et al.~\cite{GHLW15}, then we increase the space by a logarithmic factor but we can return the locations of all matches quickly.

\begin{theorem}
\label{thm:jumbled}
Given a string \(S [1..n]\) over an alphabet of size $\sigma$ and a length $k \ll n$,  with high probability in $\Oh{k n + n \sigma}$ expected time we can store \((2n \log \sigma)(1+o(1))\) bits such that later we can determine in $\Oh{k \log \sigma}$ time if a pattern of length $k$ has a jumbled match in $S$.

\end{theorem}

\section*{Acknowledgements}

Many thanks to Rayan Chikhi and the anonymous reviewers for their comments.

\bibliographystyle{splncs03}
\bibliography{mphfs}

\begin{thebibliography}{10}
\providecommand{\url}[1]{\texttt{#1}}
\providecommand{\urlprefix}{URL }

\bibitem{ACLL14}
Amir, A., Chan, T.M., Lewenstein, M., Lewenstein, N.: On hardness of jumbled
  indexing. In: Automata, Languages, and Programming, pp. 114--125. Springer
  (2014)

\bibitem{Ban12}
Bankevich, A., et~al.: {SPAdes}: A new genome assembly algorithm and its
  applications to single-cell sequencing. Journal of Computational Biology  19,
   455--477 (2012)

\bibitem{BBPV09}
Belazzougui, D., Boldi, P., Pagh, R., Vigna, S.: Monotone minimal perfect
  hashing: searching a sorted table with o (1) accesses. In: Proceedings of the
  twentieth Annual ACM-SIAM Symposium on Discrete Algorithms. pp. 785--794.
  Society for Industrial and Applied Mathematics (2009)

\bibitem{belazzougui2016bidirectional}
Belazzougui, D., Gagie, T., M{\"a}kinen, V., Previtali, M., Puglisi, S.J.:
  Bidirectional variable-order de {Bruijn} graphs. In: LATIN 2016: Theoretical
  Informatics, pp. 164--178. Springer (2016)

\bibitem{boucher2015variable}
Boucher, C., Bowe, A., Gagie, T., Puglisi, S.J., Sadakane, K.: Variable-order
  de {Bruijn} graphs. In: Data Compression Conference (DCC), 2015. pp.
  383--392. IEEE (2015)

\bibitem{bowe2012succinct}
Bowe, A., Onodera, T., Sadakane, K., Shibuya, T.: Succinct de {Bruijn} graphs.
  In: Algorithms in Bioinformatics, pp. 225--235. Springer (2012)

\bibitem{BCFL12}
Burcsi, P., Cicalese, F., Fici, G., Lipt{\'a}k, Z.: Algorithms for jumbled
  pattern matching in strings. International Journal of Foundations of Computer
  Science  23(02),  357--374 (2012)

\bibitem{wabi}
Chikhi, R., Rizk, G.: Space-efficient and exact de {Bruijn} graph
  representation based on a {Bloom} filter. Algorithm. Mol. Biol.  8(22) (2012)

\bibitem{GHLW15}
Gagie, T., Hermelin, D., Landau, G.M., Weimann, O.: Binary jumbled pattern
  matching on trees and tree-like structures. Algorithmica  73(3),  571--588
  (2015)

\bibitem{KR87}
Karp, R.M., Rabin, M.O.: Efficient randomized pattern-matching algorithms. IBM
  Journal of Research and Development  31(2),  249--260 (1987)

\bibitem{KRR13}
Kociumaka, T., Radoszewski, J., Rytter, W.: Efficient indexes for jumbled
  pattern matching with constant-sized alphabet. In: Algorithms--ESA 2013, pp.
  625--636. Springer (2013)

\bibitem{mehlhorn1982program}
Mehlhorn, K.: On the program size of perfect and universal hash functions. In:
  Foundations of Computer Science, 1982. SFCS'08. 23rd Annual Symposium on. pp.
  170--175. IEEE (1982)

\bibitem{cascading}
Salikhov, K., Sacomoto, G., Kucherov, G.: Using cascading {Bloom} filters to
  improve the memory usage for de {B}ruijn graphs. In: Algorithms in
  Bioinformatics, pp. 364--376 (2013)

\bibitem{DBLP:journals/bioinformatics/SalmelaR14}
Salmela, L., Rivals, E.: Lordec: accurate and efficient long read error
  correction. Bioinformatics  30(24),  3506--3514 (2014),
  \url{http://dx.doi.org/10.1093/bioinformatics/btu538}

\bibitem{siren2014indexing}
Sir{\'e}n, J., V{\"a}lim{\"a}ki, N., M{\"a}kinen, V.: Indexing graphs for path
  queries with applications in genome research. IEEE/ACM Transactions on
  Computational Biology and Bioinformatics (TCBB)  11(2),  375--388 (2014)

\bibitem{zerbino2008velvet}
Zerbino, D.R., Birney, E.: Velvet: algorithms for de novo short read assembly
  using de {Bruijn} graphs. Genome research  18(5),  821--829 (2008)

\end{thebibliography}

\end{document}